\documentclass[prl,twocolumn,superscriptaddress,floatfix,showpacs]{revtex4}

\usepackage{amsfonts}
\usepackage{latexsym}
\usepackage{amsthm}
\usepackage{amssymb}
\usepackage{amsmath}
\usepackage{graphicx}
\usepackage{bm}
\usepackage{color}
\usepackage[
colorlinks=true,linkcolor=red,citecolor=green,urlcolor=blue,
pdfstartview=FitH,
bookmarksopen=true,bookmarksopenlevel=0,
pagebackref=false,
pdfauthor={Kaveh Khojasteh, Daniel A. Lidar, Lorenza Viola},
pdftitle={Arbitrarily Accurate Dynamical Control in Open Quantum Systems}
pdfsubject={quantum dynamical decoupling},
pdfkeywords={quantum, dynamical decoupling, dynamically corrected gates}]{hyperref}

\begin{document}

\title{Arbitrarily Accurate Dynamical Control in Open Quantum Systems}

\author{Kaveh Khodjasteh}
\affiliation{\mbox{Department of Physics and Astronomy, Dartmouth
College, 6127 Wilder Laboratory, Hanover, NH 03755, USA}}
\author{Daniel A. Lidar}
\affiliation{\mbox{Departments of Chemistry,
Electrical Engineering, and Physics, and}
\mbox{Center for Quantum Information
Science \& Technology, University of Southern
California, Los Angeles, CA 90089, USA}}
\author{Lorenza Viola}
\affiliation{\mbox{Department of Physics and Astronomy, Dartmouth
College, 6127 Wilder Laboratory, Hanover, NH 03755, USA}}

\begin{abstract}
We show that open-loop dynamical control techniques may be used to
synthesize unitary transformations in open quantum systems in such a
way that decoherence is perturbatively compensated for to a desired
(in principle arbitrarily high) level of accuracy, which depends only
on the strength of the relevant errors and the achievable rate of
control modulation. Our constructive and fully analytical solution
employs concatenated dynamically corrected gates, and is applicable
independently of detailed knowledge of the system-environment
interactions and environment dynamics. Explicit implications for
boosting quantum gate fidelities in realistic scenarios are addressed.
\end{abstract}

\maketitle

The demand for an exquisite degree of control over the dynamics of
open quantum systems is widespread across quantum physics and
engineering, ranging from high-resolution spectroscopy and chemical
reaction control \cite{Mabuchi05}, to quantum-limited metrology
\cite{QM} and quantum information processing (QIP)
\cite{NielsenBook}. Achieving a sufficiently small `error per gate'
(EPG) is, in particular, an essential ingredient to ensuring that
fault-tolerant quantum computation is possible in spite of the
decoherence that inevitably plagues real-world devices. While
closed-loop techniques, in the form of fault-tolerant quantum error
correction (QEC) \cite{Knill-Noisy,Aliferis06}, offer thus far the
only complete prescription to meet this challenge, open-loop
\emph{dynamical QEC} is emerging as a promising alternative.  Inspired
by coherent averaging in magnetic resonance \cite{HaeberlenBook} and
exemplified in its simplest form by dynamical decoupling (DD)
\cite{DD98,CDD,UDD}, dynamical QEC aims to suppress the interaction
between the system and its \emph{quantitatively unspecified}
environment through suitable sequences of unitary
operations. Recently, DD has enabled decoherence-protected
\emph{storage} in QIP platforms as diverse as electron-nuclear systems
\cite{Morton08}, photonics qubits \cite{Vitali2008}, and trapped ions
\cite{BiercukDD}, as well as found application in suppressing
collisional decoherence in cold atoms \cite{Davidson09}.

As the gap between theory and implementations shrinks, and a growing
experimental effort is devoted to robust \emph{manipulation} of
quantum states, it is imperative that realistic constraints be
accommodated from the outset in dynamical QEC design. In practice,
control resources always entail \emph{finite} power and bandwidth,
thus precluding instantaneous (`bang-bang' \cite{DD98}) pulses. A path
toward decoherence-protected unitary operations was recently proposed
based on \emph{dynamically corrected gates} (DCGs) \cite{DCGs}.  A DCG
may be viewed as a composite quantum gate constructed from individual
(`primitive') building blocks whose errors combine non-linearly to
achieve a substantially smaller net error. If $\tau _{\text{min}}$ is
the minimum duration over which each primitive gate is effected
(`switching time') and $\Vert H_{e}\Vert $ the strength of the
error-inducing Hamiltonian, DCGs remove the effect of $H_{e}$ to the
leading (first) order, that is, the resulting EPG scales as
${\mathcal{O}}[(\tau _{\text{min}}\Vert H_{e}\Vert )^{2}]$. This
prompts the following key question: Can one make DCGs as accurate as
desired, using realistic control resources? The answer is not
obvious. Schemes capable of arbitrarily suppressing decoherence during
storage have been identified -- notably, concatenated DD \cite{CDD}
and recent optimized protocols \cite{UDD,West2009} -- but, thus far,
only in the bang-bang limit. With bounded controls, decoherence
suppression up to the second order (with leading corrections
${\mathcal{O}}[(\tau _{\text{min}}\Vert H_{e}\Vert )^{3}]$) may be
achieved by using a time-symmetrized Euler DD (EDD) protocol
\cite{Viola2003Euler} (see also \cite{RUDD}) -- however, this
procedure extends neither to generic quantum gates nor to generic open
quantum systems \cite{DCGs,Pasini2009}.

Here we show that decoherence suppression can in principle be pushed
to an order limited only by the strength of the relevant errors and
the achievable rate of control modulation. We do this by combining DCG
constructions with recursive design -- resulting in \emph{concatenated
DCGs}. While perturbative in nature, our solution is fully analytical,
laying the foundation for rigorous complexity analysis and
optimization in dynamical QEC. Not only do concatenated DCGs exist for
arbitrary finite-dimensional open quantum systems with a bounded
$\Vert H_{e}\Vert $, but they are also highly \emph{portable}, in the
sense that no quantitative knowledge of the underlying interaction
Hamiltonian is assumed, beyond its \emph{algebraic} form. Since
arbitrarily accurate open-loop compensation techniques for classical
(static) control errors are known \cite{Brown2004}, our results imply
that no fundamental limitation arises due to a quantum (dynamic) bath.
From a practical standpoint, concatenated DCGs offer the first
systematic feedback-free framework for designing quantum gates which
can achieve the arbitrarily high levels of protection against
decoherence demanded by high-fidelity quantum control, and in
particular QIP.

\emph{Control-theoretic setting.---}Let $S$ be the target quantum
system, coupled to its quantum bath $B$ via an interaction Hamiltonian
$H_{SB}$, with respective Hilbert space $\mathcal{H}_S$ and
$\mathcal{H}_B$. We assume that the total \emph{error Hamiltonian}
$H_{e}$ may be described by a joint time-independent operator of the
following form:
\begin{equation}
H_{e}=H_{S,e}+H_{SB}+H_{B}\equiv\sum_{\alpha}S_{\alpha}\otimes
B_{\alpha}, \label{eq:hehsbexp}
\end{equation}
where the contribution $H_{S,e}$ accounts for undesired `drift' terms
possibly present in the system's internal Hamiltonian,
$\{S_{\alpha}\}$ is a Hermitian basis of operators acting on
$\mathcal{H}_S$, and $B_{\alpha}$ are bounded (potentially unknown)
operators acting on $\mathcal{H}_B$.  The vector space $\Omega_{e}$
spanned by $\{S_{\alpha}\otimes B_{\alpha}\}$, with non-zero
$B_{\alpha}$, defines the \emph{error model} and is uniquely
determined by the system components $\{S_\alpha\}$ appearing in the
expansion of $H_e$ in Eq. (\ref{eq:hehsbexp}). In dynamical QEC, a
classical controller is adjoined to $S$ through a time-dependent
Hamiltonian $H_{\text{ctrl}}(t)$.  Several constraints may restrict,
in reality, the degree of control that is available through
$H_{\text{ctrl}}(t)$. In particular, we account for realistic control
modulations to be bounded in amplitude and spectral bandwidth by
requiring that $\Vert H_\text{ctrl}(t)\Vert<\infty$ and
$\tau_{\text{min}}>0$, respectively. In an ideal situation where
$H_{SB}=0=H_{S,e}$, the errors arise from imperfections in the
controller only. Our goal here is to present a proof-of-concept for
robustness against \emph{decoherence errors during gates}, thus we
ignore such contributions henceforth. We furthermore assume that
\emph{universal} control over $S$ is achievable \cite{NielsenBook}
(see \cite{DCGs} for departures from this `minimal' setting).

Consider a unitary operation (gate) $Q$ on ${S}$, which is ideally
realized by letting $H_{\text{ctrl}}(t)=H_{Q}(t)$ over an interval
$[t_{1},t_{1}+\tau ]$. We use $Q[\tau ]$ or $Q$ to denote this
implementation of $Q$ when there is no ambiguity. If $H_{e}\neq 0$,
the joint propagator is $U_{Q[\tau]}={\mathcal{T}} \exp
[-i\int_{t_{1}}^{t_{1}+\tau }(H_{Q}(t)+H_{e})dt]$ ($\hbar =1$ and
${\cal T}$ denotes time-ordering).  The effect of $H_{e}$ may be
isolated through a Hermitian `error action operator' $ E_{Q[\tau ]}$
\cite{DCGs}, where $U_{Q[\tau ]}=Q\exp (-iE_{Q[\tau ]})$.  Let the
initial states of $S$ and $B$ be $\rho_{S}^{0} =|\psi \rangle \langle
\psi |$ and $\rho _{B}$.  Ideally, the action of $Q$ would result in
$\rho _{S}^{0}(\tau )=Q\rho_{S}^{0}Q^\dagger$, while $B$ evolves
independently.  In contrast, the actual evolution is coupled and $\rho
_{S}(\tau )=\text{Tr}_{B}(U_{Q[\tau ]}\rho _{S}^{0}\otimes
\rho_{B}U_{Q[\tau ]}^{\dagger })$.  An appropriate performance measure
for control is the trace-norm distance $\Delta \equiv \Vert \rho
_{S}(\tau )-\rho _{S}^{0}(\tau )\Vert _{1}$ \cite{NielsenBook}. One
can show that $\Delta \leq \Vert \text{mod}_{B}(E_{Q[\tau ]})\Vert $,
\emph{independently of the initial state}, where
$\text{mod}_{B}(E)\equiv E-\frac{1}{\text{Tr}(I_{S})}I_{S}\otimes
\text{Tr}_{S}(E)$ is a projector that removes the pure-bath terms in
$E$ \cite{norms,CDCGLong}.  Thus, $\eta \equiv \Vert
\text{mod}_{B}(E_{Q[\tau ]})\Vert $ may be taken to quantify the
resulting EPG.

Given a desired unitary gate $Q$, our task is to synthesize a control
modulation that approximates $Q$ with an EPG scaling as
$\mathcal{O}(\tau_{\text{min}}^{\ell+1})$, for an arbitrary positive
integer $\ell$ and compatible with the stated control constraints. Any
such construction \emph{defines} an $\ell$-th order DCG,
$\text{DCG}^{[\ell]}$. ``Na\"{\i}ve'' implementations that simply
correspond to turning on, say, a constant $H_{ \text{ctrl}}(t)$ for a
duration $\tau$ yield an EPG that scales (approximately) linearly with
$\tau$ in the presence of $H_{e}$ and are included as $\ell=0$.
Primitive gates are correspondingly denoted by $\{Q^{[0]}\}$.  Free
evolution under $H_{e}$ [$H_{\text{ctrl}}(t)\equiv 0$] may be viewed
also as a zeroth order `no-operation' (\textsc{noop}) gate.  Consider
a combined gate $(Q_{1}\cdots Q_{N})[\sum \tau _{i}]$ formed as a
sequence of $N$ primitive gates $Q^{[0]}_{N}[\tau _{N}]\cdots
Q^{[0]}_{2}[\tau _{2}]Q^{[0]}_{1}[\tau_{1}]$ applied back to back.  We
may compute the total error for the combined gate using
\begin{eqnarray}
&&E_{(Q_{1}\cdots Q_{N})[\sum \tau _{i}]}=E_{Q_{1}[\tau _{1}]}+
\label{eq:phiqi-1} \\
&&+\;P_{1}^{\dagger }E_{Q_{2}[\tau_{2}]}P_{1}+\cdots +P_{N-1}^{\dagger
}E_{Q_{N}[\tau _{N}]}P_{N-1}+E_{2+}, \notag
\end{eqnarray}
where $P_{j}=Q_{j}\cdots Q_{1}$, $j=1,\ldots ,N-1$, is the `partial'
control propagator at the end of the $j$th segment \cite{MagnusNote}
and $\left\Vert E_{2+}\right\Vert =\mathcal{O}(\sum_{i,j}\left\Vert
[E_{Q_{i}[\tau _{i}]},E_{Q_{j}[\tau _{j}]}\right\Vert )$ includes the
leading second-order corrections.  Note that individual EPGs \emph{do
not simply add up, but are `modulated' by the control trajectory}: in
dynamical QEC, this provides the basic mechanism enabling the effect
of $H_{e}$ to be perturbatively canceled, without quantitative
knowledge of the bath.

\emph{Concatenated DCGs: Construction and performance bound.---} Our
strategy to achieve an arbitrary order of cancellation is to invoke a
recursive construction, that generates $(\ell +1)^{\mathrm{th}}$-order
gates using $\ell^{\mathrm{th}}$-order building blocks, close in
spirit to concatenated DD \cite{CDD}.  In what follows, given a target
gate $Q$, we shall term an {\em $\ell $-th order balance pair} as any
pair of gates $(I_{Q}^{[\ell ]},Q_{\ast }^{[\ell ]})$ whose errors
coincide up to the leading $(\ell+1)^{\text{th}}$-order, that is,
$\text{mod}_{B}(E_{I_{Q}})=\text{mod}_{B}(E_{Q_{\ast
}})+\mathcal{O}(\tau _{\text{min}}^{\ell+2})$.

The first step in the recursion requires demonstrating that, given
$Q$, a corresponding DCG$^{[1]}$ can be constructed out of the
available universal set of primitive gates.  As established in
\cite{DCGs}, this necessitates two main ingredients: (i) An EDD
protocol for generating a DCG$^{[1]}$-implementation of \textsc{noop};
(ii) A zeroth-order balance pair $(I^{[0]}_{Q}$, $Q^{[0]}_{\ast })$,
with the same leading order error $E_{Q_{\ast }}$ belonging to a
`correctable error space' $\Omega_{\mathcal{D}}\supseteq \Omega_{e}$.
In order to achieve (i), a set of unitary gates
$\mathcal{D}=\{D_{i}\}_{i=1}^{d}$ is identified, such that the map
$\Pi _{\mathcal{D}}[E]\equiv
\frac{1}{d}\sum_{i=1}^{d}D_{i}^{\dagger}ED_{i}$ `decouples' all errors
$E$ in $\Omega _{\mathcal{D}}$:
$\text{mod}_{B}(\Pi_{\mathcal{D}}[E])=0$ \cite{DD98}. In EDD
\cite{Viola2003Euler}, $\mathcal{D} $ represents a group $\mathcal{G}$
(faithfully and projectively), with order $|\mathcal{G}|=d$ and a set
of $m_\ell$ generators $\Gamma =\{F_{j}\}$. Let $G(\mathcal{G},\Gamma
)$ denote the Cayley graph of $\mathcal{G}$ with respect to $\Gamma
$. The required EDD sequence (of length $N_{1}=dm$) is constructed by
consecutively applying the generators $F_{j}$ as gates, in the order
determined by an Eulerian cycle on $G(\mathcal{G},\Gamma )$, starting
(and ending) at the identity vertex, $I_{S}$. Suppose that each
$F_{j}$ is implemented as a zeroth order gate with $E_{F_{j}}\in
\Omega _{\mathcal{D}}$. Then, using Eq.~(\ref{eq:phiqi-1}) and the
decoupling property of $ \mathcal{G}$, the net error
$\eta_{\text{EDD}} = \mathcal{O}(\text{max}_{j}\, \Vert
E_{F_{j}}\Vert^{2})=\mathcal{O}(\tau _{\text{min}}^{2})$, as desired.

In order to extend the EDD construction to a gate $Q$ other than
\textsc{noop}, some information about \emph{how} primitive gates are
implemented is required, due to a NoGo theorem for `control-oblivious'
design proved in \cite{DCGs}; balance pairs are required as ingredient
(ii) precisely for this purpose.
Once such a balance pair $(I^{[0]}_{Q}$, $Q^{[0]}_{\ast})$ is found, a
DCG$^{[1]}$ sequence for $Q$ is obtained by `augmenting' the Cayley
graph for \textsc{noop}: $I^{[0]}_{Q}$ gates are inserted in the EDD
sequence at points where the corresponding Eulerian path visits each
non-identity vertex for the first time, and finally $Q^{[0]}_{\ast}$
is applied after the last $I_{S}$ vertex. The resulting error for the
combined sequence is then given by
$\Vert\text{mod}_{B}(E_{Q^{[1]}})\Vert=
\Vert\text{mod}_{B}(E_{\text{EDD}}+\Pi_{\mathcal{D}}(E_{Q_{*}}))\Vert=0$
plus corrections given by error $\eta^{[1]}_{\text{DCG}}=
\mathcal{O}(\tau_{\text{min}}^{2})$, as desired.

Our balance pair construction ``strikes a balance'' by
\emph{stretching control profiles}.  Not only does this ensure a fully
portable recipe in terms of control inputs (note that previous DCG
constructions \cite{DCGs}, required access to sign-reversed control
profiles), but it is in fact crucial for producing balance pairs of
high-order gates. In practice, a stretched control input results in
slower gates.  More concretely, for a gate stretched by a factor $r$,
the gating Hamiltonian is ``stretched'' by a factor $r$: $
H_{Q}(t)\mapsto H_{Q}[t_{1}+r(t-t_{1})]/r$; thus the same target gate
$Q$ is approximated, but with a different EPG. The stretched gate is
symbolically indicated by $Q[r\tau]$ but all gate constructions are
still subject to the original bandwidth/power constraints. The general
recipe for balance pairs of any order $\ell$ is given in terms of
$Q^{[\ell]}$ and $Q^{-1,[\ell]}$ ($\ell^{\text{th}}$ order
implementations of $Q$ and its inverse) by the following pair of
composite gates \cite{CDCGLong}:
\begin{eqnarray}
I_{Q}^{[\ell ]}&=&Q^{-1,[\ell]} [\tau] \, Q^{[\ell]}[2^{1/(\ell+1)
}\tau ], \nonumber \\ Q_{\ast}^{[\ell ]}[\tau] &=& Q^{[\ell]}[\tau] \,
Q^{-1,[\ell]}[\tau] \, Q^{[\ell]} [\tau].
\label{balpair}
\end{eqnarray}
Note that $I_{Q}^{[\ell ]}$ implements the \textsc{noop} gate
($I_{S}$) over a duration of $\tau +2^{1/(\ell+1) }\tau $, while
$Q_{\ast }^{[\ell ]}$ implements $Q$ over $3\tau $.  Setting $\ell=0$
thus completes the construction of $1^{\mathrm{st}}$-order gates using
$0^{\mathrm{th}}$-order building blocks.

A (universal) set of gates $\{Q^{[\ell ]}[\tau _{\ell }]\}$ with EPG
$= \mathcal{O}(\tau_{\text{min}}^{\ell +1})$ can be constructed
recursively for arbitrary $\ell \geq 1$ at this point.  Let
$\Omega_{e}^{[\ell ]}$ denote the error model for all
$\ell^{\mathrm{th}}$-order gates: $E_{Q^{[\ell ]}}\in
\Omega_{e}^{[\ell ]}$ \cite{Expand}.  We can then identify the
smallest DD group $\mathcal{G}^{[\ell ]}$ that decouples $\Omega
_{e}^{[\ell ]}$, with $m_{\ell }$ group generators $\Gamma^{\lbrack
\ell ]}=\{F_{j}^{[\ell ]}\}$, and a corresponding Cayley graph
$G(\mathcal{G}^{[\ell ]},\Gamma^{\lbrack \ell ]})$. We modify this
graph by attaching self-directed edges representing $I_{Q}^{[\ell]}$
to all vertexes except $I_{S}$, and add a new vertex $Q$ by connecting
it to the $I_{S}$-vertex through an edge representing
$Q_{\ast}^{[\ell]}$. By construction, every edge in this graph
represents a DCG$^{[\ell]}$.  We now proceed as in first-order DCGs,
and implement the sequence $ Q^{[\ell+1]}[\tau _{\ell +1}]$
by following the $N_{\ell }=d_{\ell }m_{\ell }+d_{\ell }$ edges of an
Eulerian path on the modified graph for $G(\mathcal{G}^{[\ell]},\Gamma
^{\lbrack \ell ]})$, starting at $ I_{S}$ and stopping at $Q$, and
applying the corresponding $Q_{i}^{[\ell ]}$ gates back to back. If
the latter are implemented with duration $\tau_\ell$ (before
stretching), then the combined total duration $\tau_{\ell +1}$
satisfies:
\begin{equation*}
\tau _{\ell +1}=[d_{\ell }m_{\ell }+(d_{\ell }-1)(1+2^{1/(\ell
+1)})+3]\tau _{\ell }.
\end{equation*}

By iterating, we obtain a cascade of stretched primitive control
profiles of duration $\tau_\ell \lesssim (\chi_\ell)^\ell \tau_0$,
where $\chi_\ell = d_\ell(m_\ell+3)$ \cite{AdvNote1}.  Starting with
the primitive gates $[\ell =0]$ of duration $\tau _{0}$, the above
construction generates a DCG$^{[\ell ]}$ for any $Q$ and $\ell $, with
a net error upper bounded by:
\begin{equation}
\eta_{\text{DCG}}^{[\ell]} < c\, (\chi_\ell)^{\ell^2} \tau_{0}\left\Vert
H_{SB}\hspace*{-.5mm}+H_{S,e}\right\Vert \left( 4\chi_\ell \tau_{0}\left\Vert
H_{e}\right\Vert \right)^{\ell },
\label{bound}
\end{equation}
where $c={\cal O}(1)$.  While a detailed proof will be presented
elsewhere \cite{CDCGLong}, two main steps are involved. First, we
prove that for any gate $Q^{[\ell]}[\tau ]$, owing to the recursive
design of the sequence, the Magnus expansion of the error $E_{Q^{[\ell
]}}$ in terms of the toggling frame error Hamiltonian
$H_{e}(t)=U_{\text{ctrl}}^{\dagger}(t)H_{e}U_{\text{ctrl} }(t)$
contains only terms that start at ${\cal O}(\tau^{\ell+1})$ (modulo
the pure bath terms).  Second, we bound these higher order terms
(hence $\Vert\text{mod}_{B}(E_{Q^{[\ell ]}}\Vert$)) using standard
operator inequalities.

\emph{Concatenated DCGs: Analysis and applications.---} From a
practical perspective, our result above shows that concatenated DCGs
offer concrete error reduction over primitive gates and DCG$^{[1]}$:
for any fixed achievable switching time $\tau_0$, minimizing the bound
in Eq.~(\ref{bound}) yields an optimal concatenation level
$\ell_{\text{opt}} = \lfloor -\frac{1}{2} (\log_\chi (4 \Vert H_e\Vert
\tau_0) +1) \rfloor $, where $ \lfloor x\rfloor$ is the largest
integer $\leq x$, and $\chi\equiv
\chi_{\ell_{\text{opt}}}$. Substituting $\ell_{\text{opt}}$ into
Eq. (\ref{bound}) yields an EPG bound that may enable scalable QIP
even if primitive gates, whose EPG is given by $\Vert H_{SB}+
H_{S,e}\Vert \tau_0$, are above the accuracy threshold of
(non-Markovian) fault-tolerant QEC \cite{Hui}.  Ultimately, the
viability of a concatenated DCG will be dictated by system-dependent
implementation trade-offs, based on both the total gate duration and
minimum switching time.

\begin{figure}[tbp]
\begin{centering}
\includegraphics[width=3.1in]{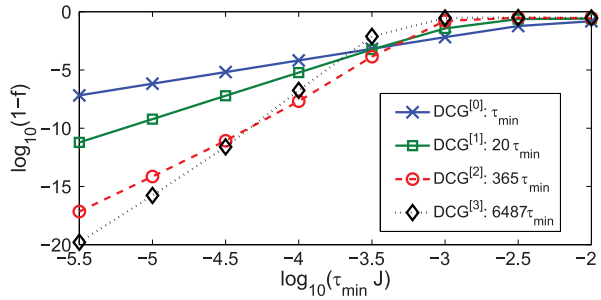} \par\end{centering}
\vspace{-.3cm}
\caption{Fidelity of increasing DCG orders for $Q=\exp [-i(\protect\pi
/3)X]$ applied to $|\protect\psi \rangle =(\left\vert 0\right\rangle
+\left\vert 1\right\rangle )/\protect\sqrt{2}$, and $\protect\rho
_{B}$ fully mixed.  The fidelity is $f
\equiv\text{Tr}\protect\sqrt{\!\protect\sqrt{\protect\rho_{a}}
{\protect\rho_{t}}\protect\sqrt{\protect\rho_{a}}}$, where
$\protect\rho_{t}$ ($\protect\rho _{a}$) is the target (actual) final
qubit state, and satisfies $1-\Delta \leq f \leq \sqrt{1-\Delta^2}$
\cite{Fuchs-distance}.  The bath spins couple to the central system
spin $\mathbf{S}$ via a Heisenberg interaction
$H_{SB}=\sum_{i=1}^{5}j_{i}\mathbf{S}\cdot \mathbf{I}^{(i)}$, with
$j_{i}$ randomly picked in $[0,J]$, $J\equiv 10$ (arbitrary
units). The bath spins $I^{(j)}$ evolve under a dipolar interaction
$H_{B}= \sum_{i,j}b_{ij}(I_{X}^{(i)}I_{X}^{(j)}+I_{Y}^{(i)}
I_{Y}^{(j)}-2I_{Z}^{(i)}I_{Z}^{(j)}) $, with $b_{ij}$ randomly picked
in $[0,10^{-2}]$.  Primitive gates were implemented using rectangular
pulse shapes to allow for numerically exact simulations. For fixed
$J$, reducing $\protect\tau _{\min }$ may be understood in terms of a
finer temporal resolution of the resulting `digitized pulse
profile'. Note that at $\log _{10}(\protect\tau_{\min }J)=-5.5$ the
fidelity for DCG$^{[3]}$ is ${13}$ orders of magnitude better than for
DCG$^{[0]}$, in spite of the gate taking nearly $6500$ times longer.
}
\label{fig:cdcg}
\end{figure}

To illustrate our general construction, consider the paradigmatic case
of a single qubit undergoing {\em arbitrary decoherence}, whereby
$\Omega _{e}^{[\ell ]}\equiv \Omega ^{(1)}=\text{span}\{\sigma
_{\alpha}\otimes B_{\alpha }\}$, and $\sigma_{\alpha }=\{I,X,Y,Z\}$
are the identity and the Pauli matrices. The corresponding DD group
$\mathcal{G}=\mathbb{Z}_{2}\times \mathbb{Z}_{2}$ is (projectively)
represented as $\{I,X,Y,Z\}$, and is generated by, \emph{e.g.},
$\Gamma =\{X,Y\}$. EDD is given by $XYXYYXYX$
\cite{Viola2003Euler}. Starting with a $Q^{[\ell ]}[\tau_{\ell }]$
gate, we have:
\begin{equation*}
Q^{[\ell +1]}=Q_{\ast }^{[\ell ]}X^{[\ell ]}Y^{[\ell ]}X^{[\ell ]}Y^{[\ell
]}Y^{[\ell ]}I_{Q}^{[\ell ]}X^{[\ell ]}I_{Q}^{[\ell ]}Y^{[\ell
]}I_{Q}^{[\ell ]}X^{[\ell ]},
\end{equation*}
in which $I_{Q}^{[\ell ]}$ and $Q_{\ast }^{[\ell ]}$ are defined in
Eq. (\ref{balpair}), and are themselves combinations of stretched
$Q^{[\ell ]}$ and $Q^{-1,[\ell]}$.  The length of the combined new
sequence is given by $\tau _{\ell +1}=[14+3\times 2^{1/(\ell +1)}]\tau
_{\ell }\le20\tau _{\ell }\equiv \chi \tau _{\ell }$.  Representative
simulation results are presented in Fig.~\ref{fig:cdcg}. Parameters
have been chosen to mimic a high-quality silicon (Si) quantum dot,
where the electron spin qubit undergoes hyperfine-induced decoherence
due to a fraction $\approx 1$ ppm of non-zero spin $^{29}$Si nuclei
(about one order of magnitude larger than currently achieved
isotopically purified Si \cite{Tsubouchi}).  Consistent with the
perturbative nature of the error cancellation in dynamical QEC, the
improvement due to the recursive design is manifested in the
increasing slopes associated with higher concatenation levels once the
gating time $\tau_{\min }$ is sufficiently short.  Concatenated DCGs
may also prove instrumental in reducing gating errors on a recently
proposed \emph{logical} qubit encoded in the singlet/triplet spin
manifold of a Si double quantum dot \cite{Levy2009}, in particular to
protect exchange-based logical $Z$ rotations against magnetic field
noise.  Finally, we expect the same control sequences to be effective
for non-Markovian decoherence induced by unbounded environments with a
sufficiently `hard' spectral cutoff \cite{UDD}.

\emph{Conclusion.---}We have presented a general constructive solution
to the problem of generating arbitrarily accurate quantum gates with
finite control resources in an open-loop setting. In addition to
settling a fundamental question, our results point to several venues
for further investigation. On the theory side, improved constructions
should incorporate more realistic \emph{local} baths and combinatorial
Eulerian design \cite{WocjanEOA}. From an implementation perspective,
making contact with optimal-control formulations \cite{opengrape09}
may allow to boost efficiency in experimentally available control
platforms.

K.~K. and L.~V. gratefully acknowledge insightful discussions with Winton
Brown. This material is based upon work supported by the NSF under Grants
No. PHY-0555417 and PHY-0903727 (to LV), and CCF-726439 and PHY-803304 (to
DAL). DAL thanks IQI-Caltech, where part of his work was done.


\begin{thebibliography}{28}
\bibitem[{nor()}]{norms}
\bibinfo{note}{$\|X\|_1\equiv {\rm Tr}\sqrt{X^\dagger X}$ and $\|X\|$ is the
  standard operator norm (largest eigenvalue of $\sqrt{X^\dagger X}$). ${\rm
  Tr}_X$ denotes partial trace over the bath ($X=B$) or system ($S=X$).}
\bibitem[{Mag()}]{MagnusNote}
\bibinfo{note}{Formally, Eq.~(\ref{eq:phiqi-1}) is a truncation of the Magnus
  expansion and is valid as long as
  $\sum_{i}\Vert\text{mod}_{B}E_{Q_{i}[\tau_{i}]}\Vert<\pi$, see
  \cite{HaeberlenBook,DCGs}.}
\bibitem{Expand} Note that similarly to $H_{e}$, $E_{Q^{[\ell]}}$ may
be expanded as $E_{Q^{[\ell]}}=\sum S_{\alpha }\otimes B_{Q,\alpha }$,
where $B_{Q,\alpha }$ are operators acting on $\mathcal{H}_{B}$ that
depend on both $\{B_{\alpha }\}$ and the control sequence.
\bibitem[{Adv()}]{AdvNote1}
\bibinfo{note}{If $D=\text{dim}({\mathcal H}_S)$, in the worst case ${\mathcal
  G}^{[\ell]}$ may be identified with the `adversarial DD group' that achieves
  maximal decoupling on $S$, ${\mathcal G}_{\text{adv}}\simeq
  \mathbb{Z}_{D}\times \mathbb{Z}_{D}$, under an irreducible unitary
  representation in ${\mathcal H}_S$. Thus, $d_\ell \leq D^2$ and $m_\ell = 2$,
  for all $\ell$ \cite{DD98,WocjanEOA}. For $n$ qubits, the adversarial error
  model $\Omega_e^{(n)}$ is spanned by all combinations of up to $n$ Pauli
  matrices, and ${\mathcal G}_{\text{adv}} \simeq (\mathbb{Z}_{2}\times
  \mathbb{Z}_{2})^{\times n}$. Fixing $\Omega_e^{[\ell]}=\Omega^{(n)}_e$ yields
  worst-case CDCGs of length exponential in $n$.}
\end{thebibliography}
\end{document}